\documentclass[prl,aps,twocolumn,showpacs,superscriptaddress,twoside]{revtex4-1}
\usepackage{graphicx}
\usepackage{color}

\begin{document}
\title{First experimental realization of the Dirac oscillator}
\author{J. A. Franco-Villafa\~ne}
\affiliation{Instituto de Ciencias F\'isicas, Universidad Nacional Aut\'onoma de M\'exico, Av. Universidad s/n, 62210 Cuernavaca, M\'exico}
\author{E. Sadurn\'i}
\affiliation{Instituto de F\'isica, Benem\'erita Universidad Aut\'onoma de Puebla, Apartado Postal J-48, 72570 Puebla, M\'exico}
\author{S. Barkhofen}
\affiliation{Fachbereich Physik, Philipps-Universit\"{a}t Marburg, Renthof 5, 35032 Marburg, Germany}
\author{U. Kuhl}
\affiliation{Laboratoire de Physique de la Mati\`ere Condens\'ee, Universit\'e de Nice-Sophia Antipolis, CNRS, UMR 7336 Parc Valrose - 06108 Nice, France}
\author{F. Mortessagne}
\affiliation{Laboratoire de Physique de la Mati\`ere Condens\'ee, Universit\'e de Nice-Sophia Antipolis, CNRS, UMR 7336 Parc Valrose - 06108 Nice, France}
\author{T. H. Seligman}
\affiliation{Instituto de Ciencias F\'isicas, Universidad Nacional Aut\'onoma de M\'exico, Av. Universidad s/n, 62210 Cuernavaca, M\'exico}
\affiliation{Centro Internacional de Ciencias, 62210 Cuernavaca, M\'exico}


\begin{abstract}
We present the first experimental microwave realization of the one-dimensional Dirac oscillator, a paradigm in exactly solvable relativistic systems. The experiment relies on a relation of the Dirac oscillator to a corresponding tight-binding system. This tight-binding system is implemented as a microwave system by a chain of coupled dielectric disks, where the coupling is evanescent and can be adjusted appropriately. The resonances of the finite microwave system yields the spectrum of the one-dimensional Dirac oscillator with and without mass term. The flexibility of the experimental set-up allows the implementation of other one-dimensional Dirac type equations.
\end{abstract}


\maketitle

The relativistic version of the harmonic oscillator has been touched upon occasionally \cite{ito67,coo71}, but became a widely used model for relativistic equations with the appearance of the seminal paper \cite{mos89}. Originally it was known as the Dirac oscillator (DO) \cite{sad10b,tor10}. Indeed since then the number of papers using this model has increased rapidly, mainly in mathematical physics \cite{ham12,lu11,zar10,cha10,ber08,kul07,cas06,alh06,que05,ho04,lis04,alh01a,alh01b,ben90}, but also in nuclear physics \cite{mun12,gri12a,fae05}, subnuclear physics \cite{wa12,rom11} and quantum optics \cite{dod02,ber07a,lam07,lon10}. In mathematical physics it has become the paradigm for the construction of covariant quantum models with some well determined non-relativistic limit, but has also attracted much attention in the environment of exactly solvable models and symmetries; it is amusing to mention that even the Higgs symmetry has been considered in this context \cite{zha09}.

While this model is a paradigm of mathematical physics, it does not describe a known physical system, as is the case for the Dirac equation for the hydrogen atom. Thus an experimental realization by other means is highly desirable. There are two proposals to realize analogue experiments. One in the realm of quantum optics \cite{ber07a,lam07,lon10} and the other one using a classical microwave setup \cite{sad10a}. In this paper we shall present a microwave realization for the 1-D DO. Beyond its intrinsic interest, the experiment is also a starting point for further experimental exploration of Dirac-like equations.

We will mainly follow the proposition of Ref.~\cite{sad10a} but use a slightly different mechanism to appropriately take into account the finiteness of the experimental system. The experimental idea is based on a mapping of the DO to a tight-binding model with dimers. In this model it is important that only nearest neighbor interactions are present. It consists of a chain of coupled disks with a high index of refraction sandwiched between two metallic plates. The coupling constants between the disks have to be adjusted properly, to obtain a spectrum which is equivalent to the DO spectrum. This set-up has been used to investigate the Dirac points \cite{kuh10a}, disorder effects \cite{bar13a} and topological transitions in graphene \cite{bel13}. We start with a short introduction to the DO and its relation to a tight-binding Hamiltonian with nearest neighbor coupling only. Thereafter we introduce the experimental setup and present the experimental results.

{\it Dirac oscillator.-\ } The system that we now call the DO was proposed more than 20 years ago \cite{mos89,mos90a,mos91,mos92,mos96}, and its properties and possible applications have been studied extensively. The original formulation was presented in Hamiltonian form.  Covariance was easily achieved and the physicality of such a system could be attained by means of a Pauli coupling \cite{mos89}. In the present paper we will use the $1+1$ dimensional version of the Dirac oscillator \cite{sad10b}, which can be treated analogously and yields a two-component spinless structure.

The system in question can be conceived in its simplest form by writing the corresponding Hamiltonian as a function of the spectrum generating algebra. Let $a, a^{\dagger}$ be the ladder operators of a non-relativistic harmonic oscillator and $\sigma_{\pm}= \sigma_x \pm i \sigma_y$ the creation and annihilation operators of spin $1/2$ in terms of Pauli matrices. The $1+1$ dimensional DO Hamiltonian is
\begin{equation}\label{eq:HSigma}
H= \sigma_{+} a + \sigma_{-} a^{\dagger} + \mu \sigma_z,
\end{equation}
where the spectrum is given by
\begin{equation}\label{eq:eps_nwithm}
\epsilon_{\pm, n} = \pm \sqrt{n + \mu^2}
\end{equation}
where the sign denotes particles and antiparticles. The dimensionless commutator $[a,a^{\dagger}]=1$ ensures, that for a particle of mass $m$ and an oscillator of frequency $\omega$, we have $\mu = \sqrt{mc^2 / \hbar \omega}$. Thus in the appropriate units, $\mu$ gives the mass of the particle directly, and the time variable scales as $t \mapsto \sqrt{\omega m c^2/ \hbar} t$. In a certain limit, the DO can also be reduced to the Weyl equation with a linear potential, thus one may call this system even a Weyl oscillator. Taking $m\to 0$ leaving $m \omega=$~constant~$\neq 0$ leads to $\mu \rightarrow 0$ and the following massless Hamiltonian
\begin{equation}
H= \sigma_{+} a + \sigma_{-} a^{\dagger},
\label{2}
\end{equation}
with the simplified spectrum
\begin{equation}\label{eq:eps_n}
\epsilon_n = \pm \sqrt{n}.
\end{equation}
Note here that for $n=0$ there is a double degeneracy, where one of the states is given by $|-,0\rangle$. Both Hamiltonians (\ref{eq:eps_nwithm}) and (\ref{2}) can be described within a tight-binding model with dimers.

\begin{figure}
 \centering
 \includegraphics[width=.95\columnwidth]{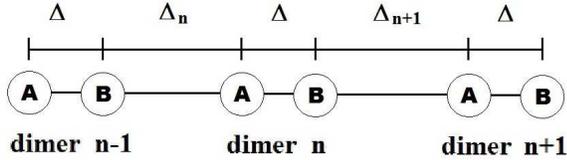}
 \caption{\label{fig:dimer}
 A chain of resonators in dimeric configuration, with atoms of type A and B. The index $n$ stands for the dimer number, $\Delta$ is the coupling between elements of the same dimer (intra-dimer coupling, kept constant throughout the array) and $\Delta_n$ is the coupling between the right end of dimer $n$ and the left end of dimer $n+1$ (inter-dimer coupling). Have in mind that here we present the couplings only, later on $\Delta\le\Delta_n$, which means that in the experiment the intra-dimer distance is the largest distance in the chain.}
\end{figure}

{\it Description as a tight-binding model.-\ } The eigenvalue problem resulting from the Hamiltonian (\ref{eq:HSigma}) can be written as two coupled tight-binding equations of the form
\begin{eqnarray} \label{3}
\sqrt{n+1} \psi^{-}_{n+1} + \mu \psi_n^{+} &=& \epsilon_n \psi_n^{+}\\
\sqrt{n} \psi^{+}_{n-1}  - \mu \psi_n^{-} &=& \epsilon_n \psi_n^{-}
\label{4}
\end{eqnarray}
where $\psi_{n}^{\pm}$ is the atomic wavefunction of the $n$-th dimer and the superscripts $+$ and $-$ indicate sites of type A and B (see also fig.~\ref{fig:dimer}). In our previous work \cite{sad10a} we have established that this model can be emulated in a one-dimensional chain with nearest neighbor interactions where the spin ($\pm$ superscripts in the equations above) can be represented by A and B sites in a linear chain. By defining the new operators
\begin{equation}
b = \Delta(1 + a), \quad b^{\dagger} = \Delta(1 + a^{\dagger}),
\label{5}
\end{equation}
the Hamiltonian for a tight-binding chain of two species can be written as
\begin{equation}
H_{\mathrm{chain}} =  \sigma_{+} b + \sigma_{-} b^{\dagger} +  \mu \sigma_z
\label{eq:HChain}
\end{equation}
and $\mu$ is now the energy difference between the resonances sitting upon A and B sites giving rise to a spectral gap. The constant $\Delta$ is nothing else than the coupling between two sites, and the spectrum of the system can be extracted by virtue of the algebraic relation $[b,b^{\dagger}]=\Delta^2$. As before, we have
\begin{equation}\label{eq:eps_nDelta}
\epsilon_n = \pm \sqrt{\Delta^2 n + \mu^2}
\end{equation}
The map between the DO and the coupled linear chain of two species is therefore quite natural. Finally we can see that the resulting array comprises dimers $AB$, e.g. sites $A$ is always equally coupled to site $B$ by $\Delta$ independently on $n$, whereas the coupling between the dimers $\Delta_n$ has to follow a specific law derived below. The requirements for a realization of $b, b^{\dagger}$, on the other hand, introduce the following restrictions: For the inter-dimer coupling $\Delta_{n} = \Delta \sqrt{n}$.

{\it An appropriate cut-off.-\ }
Till now we assumed a semi-infinite array which terminates at one end with the value $\Delta_0 = 0$ (no more dimers to the left). Therefore, couplings of the form $\sqrt{n}$ range from $0$ to $\lim_{n \rightarrow \infty} \Delta_n = \infty$. However, experimentally accessible couplings always have an upper limit $\Delta_{\mathrm{sup}}$ determined by the physical situation and such a restriction introduces a natural cut-off in the array by means of the relation $\Delta_{\mathrm{sup}} = \Delta \sqrt{n_{\mathrm{sup}}}$. Thus we arrive at a finite chain with a total number $2n_{\mathrm{sup}}$ of sites.

A previously proposed finite realization \cite{sad10a}, although well conceived for the infinite case, did not take into account cut-off effects appropriately. Any configurations of type $b = \Delta a + \delta$ for arbitrary $\delta$ fulfills the algebraic relations, but to keep edge effects small $\delta$ must be smaller than any other coupling in the system. Choosing $\delta = \Delta$ is sufficient for this purpose. The preferred tight-binding models are such that the successive couplings are increased till a maximal coupling is reached, which is in contrast with the previous proposition.

{\it The generation of mass.-\ }
Our scheme so far contemplates the appearance of a spectral gap corresponding to a finite mass in the DO to result from an inherent asymmetry within the dimer, i.e. A and B have different eigenenergies. This produces the term $\mu \sigma_z$ in the Hamiltonian [eq.~(\ref{eq:HSigma})]. However, in practice an alternate option to generate a gap occurs due to finite size effects: We choose the smallest inter-dimer coupling to be slightly smaller, rather than equal to the intra-dimer coupling, i.e. $\Delta' \gtrsim \Delta_{\mathrm{min}}$.
Numerical inspection of the tight-binding model shows that, while a gap opens, the effects on the relative position of the eigenvalues on both sides of the gap have finite size errors similar to the ones in the gapless case. Note that the gap depends on the number of sites and vanishes as this number goes to infinity; therefore a large array will not describe a DO with mass, in compliance with the chiral symmetry of the system \cite{sem06}. Yet for finite sizes we do get the desired spectrum and we can make appropriate aproximations or numerical calculations in the tight-binding model to explain this satisfactorily. Nevertheless we suggest to adapt the gap size to the experimental one rather than to get it from a tight-binding model as there will always remain discrepancies between the model and the experiment, which are entirely unrelated to the DO. If we wish to take advantage of this finite size effect we should thus replace eq.~(\ref{eq:HChain}) by
\begin{equation}\label{eq:eps_nDeltaExp}
\epsilon_n = \pm \sqrt{\Delta^2 n + \mu_{\mathrm{exp}}^2},
\end{equation}
where $\mu_{\mathrm{exp}}$ refers to the parameter determined by the experiment.

\begin{figure}
  \includegraphics[width=.95\columnwidth]{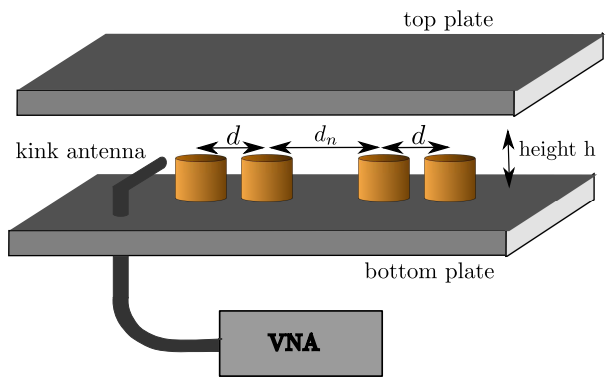}
  \caption{\label{fig:Experimentalsetup}
  Disks with a high index of refraction are placed between two metallic plates as a chain with center to center distances $d_n$ or $d$. The microwaves are induced by a vector network analyzer (VNA) via a microwave cable and a kink antenna.}
\end{figure}

{\it Experimental results.-\ }
For the experimental realization of the DO we use the techniques that have been developed to investigate the band structure of graphene \cite{kuh10a,bar13a,bel13}. The realization of the DO is achieved as tight-binding system with nearest neighbor coupling and small higher order ones. A set of identical dielectric cylindrical disks (5\,mm height, 4\,mm radius and a refractive index of about 6) is placed between two metallic plates (see fig.~\ref{fig:Experimentalsetup}). Close to one disk we placed a kink antenna connected to a vectorial network analyzer allowing to excite both transverse magnetic (TM) and transverse electric (TE) modes. The individual disks have an isolated TE resonance at 6.65\,GHz. Restricting our investigation to frequencies around this value, where each disk contributes only one resonance. The electromagnetic field for this TE mode is mostly confined within the disks and spreads evanescently outside. A sketch of the experimental setup is shown in fig.~\ref{fig:Experimentalsetup} and a detailed description is presented in ref.~\cite{bar13a}.

\begin{figure}
 \centering
 \includegraphics[width=.95\columnwidth]{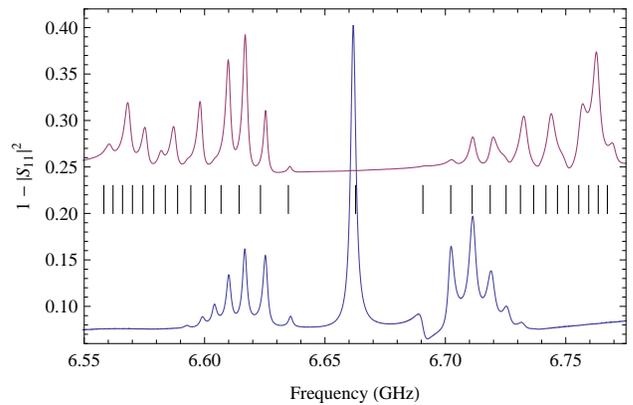}
 \caption{\label{fig:spectrum}
 Reflection spectra of a Dirac oscillator without mass for a 15 dimer chain with minimal dimer distance of $d_\mathrm{min}=13$~mm. The upper spectrum (up-shifted) is measured at the 15th disk whereas the lower is measured at 3rd disk. The vertical lines indicate the predicted resonance positions from eq.~(\ref{eq:nu_n}) with $\Delta=0.023$\,GHz.}
\end{figure}

In contrast to refs.~\cite{kuh10a,bar13a} we adjusted the height between the two plates to $h$=13\,mm, to reduce the higher order neighbor couplings. The coupling parameter $\Delta$ between two adjacent disks depends on the distance between centers of the disks $d$ and can be given in terms of a modified Bessel function $|K_0|^2$, as described in~\cite{kuh10a,bar13a}. Thus, by changing the distance between disks, one changes the inter-disk couplings and obtains the 1D-DO.

\begin{figure}
  \includegraphics[width=.95\columnwidth, height=3.5in]{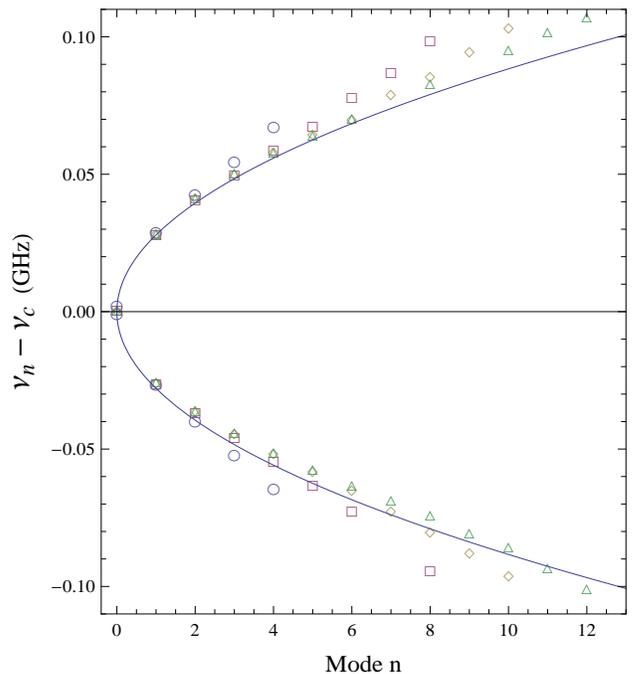}
  \caption{\label{fig:DO_13mm}
  Dirac oscillator without mass for an intra-dimer distance of $13$~mm. The continuous curve corresponds to the analytical prediction (\ref{eq:nu_n}). The symbols correspond to different numbers of dimers: 6 dimers (circles), 9 dimers (squares), 12 dimers (diamonds), and 15 dimers (triangles).}
\end{figure}

For the sake of simplicity let us assume an exponential law which is a good approximation of the coupling in terms of the distance in the range of interest. Then the distances $d_n$ between the dimers in the massless case are given by
\begin{equation}\label{eq:DistanceDelta}
d_n = -\frac{1}{\gamma} \ln \left(\frac{\Delta_n}{\Delta_K}\right)= -\frac{1}{\gamma} \ln \left(\frac{\Delta\sqrt{n}}{\Delta_K}\right).
\end{equation}
The intra-dimer distance is $d$ and we chose $d_1=d$. The distances between the dimers are decreasing monotonically, thus the smallest possible distance $d_{\mathrm{inf}}$ determined by the diameter of the disks $d_D$ defines the largest possible coupling $\Delta_{\mathrm{sup}}$ and the largest allowable number of dimers $n_{\mathrm{sup}}$ giving the largest admissible size of our dimer chain. The number of energy levels is therefore equal to $2n_{\mathrm{sup}}$. The eigenfrequencies for the 1D-DO without mass is then given by
\begin{equation}\label{eq:nu_n}
 \nu_n=\nu_c \pm\Delta\sqrt{n},
\end{equation}
where $\nu_c$ is the eigenfrequency of a single disk. We used an intra-dimer distance $d$ of $13$, $14$ and $15$~mm and chains of $12$, $18$, $24$ and $30$ disks. In fig.~\ref{fig:spectrum} we show the reflection spectra for $d$=13\,mm and 30 disks for two different antenna positions. The height of the resonances depends on the antenna site, as it is proportional to the intensity of the wavefunction at the disk. By measuring at different sites it is possible to extract all resonance positions. The vertical lines correspond to the theoretical predictions and a good agreement is found. Deviations increase at the edges of the spectrum, as designed by the choice of the cutoff.

We now investigate the dependence on the chain length. The measured eigenfrequencies as a function of the mode number is shown in fig.~\ref{fig:DO_13mm}. The continuous curve corresponds to the analytical prediction (\ref{eq:nu_n}). As the number of dimers increases, we find that the low levels are best reproduced by the theoretical curve (\ref{eq:nu_n}) and the point of departure from theory moves further away from the center of the spectrum as the number of dimers increases. For dimer distances of $14$ and $15$~mm we got similar results. Thus we experimentally measured the spectrum of the Dirac oscillator without mass in a finite approximation.

\begin{figure}
  \includegraphics[width=.95\columnwidth]{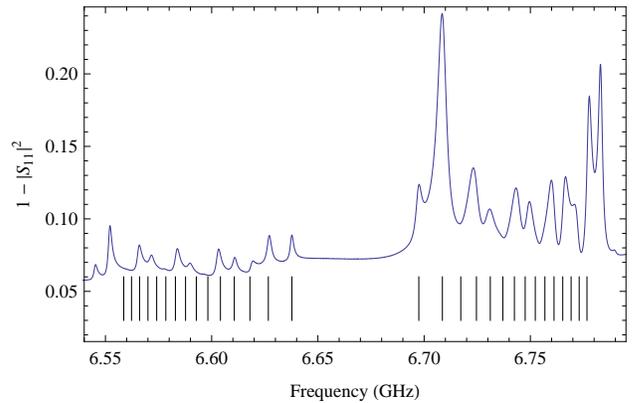}
  \caption{\label{fig:DO13mmN30Mass}
  Reflection spectrum  of a Dirac oscillator with mass for a 15 dimer chain, where the intra-dimer distance of $10$~mm and the inter-dimer distance of $d_\mathrm{min}=10.81$~mm. The vertical lines indicate the predicted resonance positions from eq.~(\ref{eq:nu_n}) with $\mu=1.066$\,GHz and $\Delta=0.028$\,GHz.}
\end{figure}

\begin{figure}
  \includegraphics[width=.95\columnwidth, height=3.5in]{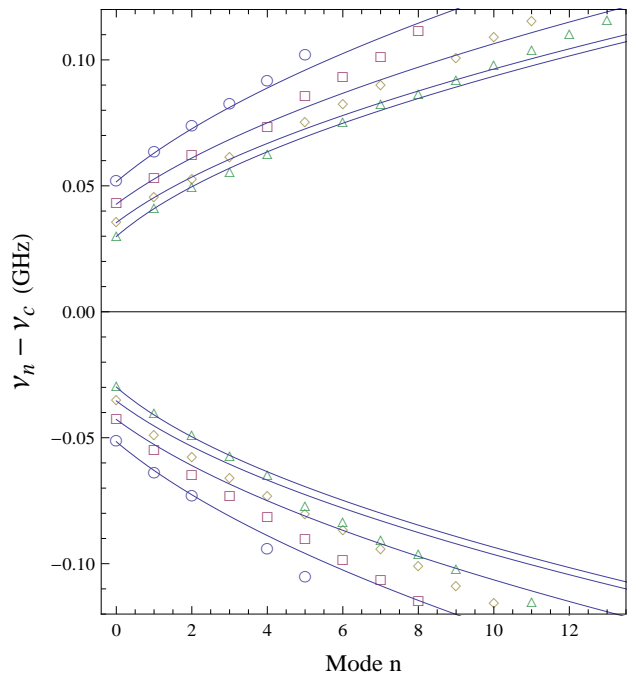}
  \caption{\label{fig:DO13mmNVaryMass}
  Dirac oscillator with mass for different number of dimers, where the last dimers are removed. The intra-dimer distance is $d$=10\,mm. Displayed are 6 ($d_\mathrm{min}\approx$12.32\,mm, Circles), 9 ($d_\mathrm{min}\approx$11.61\,mm, squares), 12 ($d_\mathrm{min}\approx$11.15\,mm, diamonds), and 15 ($d_\mathrm{min} \approx$10.81\,mm, triangles). Additionally the corresponding theoretical curves resulting from (\ref{eq:eps_nDelta}) are plotted as solid lines.
  }
\end{figure}

As we only have disks of the same type, meaning having approximately the same resonance frequency, we cannot directly generate the 1D-DO with mass as originally introduced. But, as mentioned above, for a finite chain one can introduce a mass term by setting the intra-dimer coupling $\Delta'$ larger than the smallest coupling between the dimers $\Delta_\mathrm{min}$. Thus we only have to set the intra-dimer distance $d$ to be smaller than the maximal inter-dimer distance $d_{1}$. We used a chain of 15 dimers with an initial inter-dimer distance $d_1$=15\,mm and a smallest inter-dimer distance $d_{14}=d_\mathrm{min} \approx$10.81. As intra-dimer distances $d$ we choose 10, 11, and 12\,mm. In fig.~\ref{fig:DO13mmN30Mass} we present the reflection spectra and the theoretical prediction (eq.~(\ref{eq:eps_nwithm})) for $d$=10\,mm. We observe the expected gap at the center and find a good agreement for the resonances close to the gap. Again the outer resonances show larger deviations. Next we removed step by step the last dimer, thus increasing the minimal inter-dimer distance $d_{\mathrm{min}}$ starting with $d_{\mathrm{inf}}$.

In fig.~\ref{fig:DO13mmNVaryMass} the resonances for different chain lengths are shown. We observe a good agreement for the upper spectrum with eq.~(\ref{eq:eps_nwithm}). The two bands behave slightly differently, especially their width is different, due to the second nearest neighbor couplings, as was also observed in square and graphene lattices \cite{arXbel13}. Furthermore the gap is observed to increase monotonically with $d_\mathrm{min}$.

In conclusion, we have experimentally realized the 1-D DO based on the correspondence of the DO to a tight-binding model. Within this model effects of finite size are small at the center of the spectrum. Furthermore, we have produced a gap in the spectrum which can be interpreted as the mass of the fermion. This was done by a distortion that applies only to finite arrays, as the infinite limit of the system makes such a gap vanish. We hope for the future to investigate wavefunctions and pulse propagation as well. Both are accessible for the setup if one uses a movable antenna with an additionally fixed antenna and measure the transmission~\cite{bel13,arXbel13}. Additionally we would like to realize a 2D-DO as mentioned in Ref.~\cite{sad10a}. The model assumes a logarithmically deformed hexagonal lattice with only nearest neighbor couplings. To respect this coupling condition a realization of the 2-D DO is not possible with our distance-coupling relation. However, microwave graphs seems to be a promising candidate \cite{hul04,hul05a}.

\section{Acknowledgments}

T. H. S., J. A. F. V. and S. B. thank the University of Nice for the hospitality during several long term visits at the LPMC. T. H. S. and J. A. F. V. to CONACyT Project Number 79613, PAPIIT-UNAM project number IG101113 and PAEP-UNAM for financial supports. E. S. acknowledges support from PROMEP project No. 103.5/12/4367.

\end{document}